\documentclass[aps,superscriptaddress,eqsecnum,nofootinbib,showpacs,preprintnumbers]{revtex4}
\usepackage{graphicx,epsfig}
\usepackage{amsmath}
\usepackage {amssymb}

\newcommand{\be}{\begin{eqnarray}}
\newcommand{\ee}{\end{eqnarray}}
\newcommand{\bea}{\begin{eqnarray}}
\newcommand{\nn}{\nonumber}
\newcommand{\eea}{\end{eqnarray}}

\def\del{\partial}
\def\a{\alpha}
\def\b{\beta}
\def\g{\gamma}

\def\e{\eta}
\def\la{\lambda}
\def\La{\Lambda}
\def\k{\kappa}
\def\m{\mu}
\def\n{\nu}

\def\p{\pi}

\def\z{\zeta}

\begin{document}

\title{Black Holes in Bi-scalar Extensions of Horndeski Theories}
\author{Christos Charmousis}
\email{christos.charmousis@th.u-psud.fr}
\affiliation{Laboratoire de Physique Th\'eorique (LPT), Univ. Paris-Sud, CNRS UMR 8627, F-91405 Orsay, France}
\affiliation{Laboratoire de Math\'ematiques et Physique Th\'eorique (LMPT), CNRS UMR 6083, Universit\'e Francois Rabelais-Tours, France}
\author{Theodoros Kolyvaris}
\email{theodoros.kolyvaris@ucv.cl} \affiliation{Instituto de
F\'{i}sica, Pontificia Universidad Cat\'olica de Valpara\'{i}so,
Casilla 4950, Valpara\'{i}so, Chile.}
\author{Eleftherios Papantonopoulos}
\email{lpapa@central.ntua.gr} \affiliation{Department of
Physics, National Technical University of Athens, Zografou Campus
GR 157 73, Athens, Greece.}
\author{Minas Tsoukalas}
\email{minasts@cecs.cl} \affiliation{Centro de Estudios
Cient\' ificos, Casilla 1469, Valdivia, Chile}

\date{\today}

\begin{abstract}

\end{abstract}

\date{\today}

\begin{abstract}


We study certain bi-scalar-tensor theories emanating from
conformal symmetry requirements of Horndeski's  four-dimensional
action. The former scalar is a Galileon with shift symmetry
whereas the latter scalar is adjusted to have a higher order
conformal coupling. Employing techniques from local Weyl geometry
certain Galileon higher order terms are thus constructed to be
conformally invariant. The combined shift and partial conformal
symmetry of the action, allow us to construct exact black hole
solutions. The black holes  initially found are of planar horizon
geometry embedded in anti de Sitter space and can accommodate
electric charge. The conformally coupled scalar comes with an
additional independent charge and it is well-defined on the
horizon whereas additional regularity of the Galileon field  is
achieved allowing for time dependence. Guided by our results in
adS space-time we then consider a higher order version of the BBMB
action and construct asymptotically flat, regular, hairy black
holes. The addition of the Galileon field is seen to cure the BBMB
scalar horizon singularity while allowing for the presence of
primary scalar hair seen as an independent integration constant
along-side the mass of the black hole.

\end{abstract}

\maketitle

\section{Introduction}


There has been a renewed interest in modified gravity theories due
to cosmological observations pointing towards an accelerating
Universe \cite{Amanullah:2010vv,Hinshaw:2012aka}. Gravity theories
resulting from the Horndeski Lagrangian \cite{horny} belong to a
general class of scalar-tensor theories which have been under
intense investigation recently. Apart from their mathematical
generality, there are two special reasons why Hornedski theories
are attractive. First of all they are consistent and technically
manageable. This stems from the fact that they give second-order
field equations. Secondly, a subset of these scalar-tensor theories of
modified gravity share a classical Galilean symmetry
 around
flat space-time (Galileon theories) or around curved space-time
(Generalized Galileon theories)   \cite{Nicolis:2008in,
Deffayet:2009wt, Deffayet:2009mn}.

In four dimensions, the most general Galileon theory with second order
field equations was given a long time ago by Horndeski  \cite{horny},
and can be written in the simpler form presented in
\cite{Deffayet:2011gz},
 \begin{equation}\label{horny}
S_\text{Horndeski}[\chi, g]=\int d^4 x \sqrt{-g} \left[K(\chi,
X)-G_3(\chi, X) {\cal E}_1 \right.
\\ \left.
+G_4 (\chi, X)R+G_{4, X} {\cal E}_2 +G_5(\chi, X)
G_{\mu\nu}\nabla^\mu \nabla^\nu \chi-\frac{G_{5, X}}{6} {\cal E}_3
\right]
 \end{equation}
where $X=-\frac{1}{2} (\nabla \chi)^2$, ${\cal E}_n=n!
\nabla_{[\mu_1}\nabla^{\mu_1} \chi\cdots
\nabla_{\mu_n]}\nabla^{\mu_n} \chi$ and commas denote
differentiation i.e. $G_{4,X}=\frac{\del G_4}{\del X}$. Note that
the functions $G_i$ appearing in (\ref{horny}) are in general
functions of the scalar field $\chi$ and its kinetic term $X$. In
a nutshell Horndeski or Galileon theory is the most general
scalar-tensor theory with second order field equations in
four-dimensional space-time.

Few local solutions are known, due to the complexity of
this higher order theory (\ref{horny}). One way of approach is via known black hole solutions of higher dimensional metric theories.
It has been known since a long time, that Lovelock
theory (for a review see \cite{lovelock}) yields several  Gallileon terms by
Kaluza-Klein compactifications \cite{MuellerHoissen}, \cite{VanAcoleyen:2011mj}.
Following this path, four dimensional analytic black hole solutions of Horndeski theories
were first found by Kaluza-Klein reduction of higher dimensional
Lovelock theory \cite{Charmousis:2012dw}.
Interestingly the higher order solutions cloak naked singularities
of lower order Einstein-dilaton theories, by introducing a novel
event horizon. They are however not asymptotically flat. At about the same time a nice no hair argument was introduced concerning asymptotically flat solutions
for Gallileons \cite{Hui:2012qt}.
There it was argued, under some generic hypotheses, that static, spherically
symmetric black hole solutions for the gravity-galileon coupled
system could not sustain primary scalar hair,  for vanishing
boundary condition at infinity (see also \cite{Germani:2011bc}). To
prove this the shift-symmetry of the galileon action and the
regularity of diffeomorphic invariant quantities at the horizon
was used. Ways to circumvent the no hair argument have been discussed in \cite{Babichev:2013cya},\cite{Sotiriou:2013qea} and we will exploit these here explicitly.

One of the elegant terms appearing in  the Horndeski Lagrangian is the
derivative coupling of the scalar field to the Einstein tensor
\be\label{EGBscalaraction}
   I=\int  d^4x\sqrt{-g}\left[ \frac{R}{16\pi G}-(g^{\mu\nu}-G(\chi)
   G^{\mu\nu})\nabla_\mu\chi \nabla_\nu\chi  \right]~.
\ee pictured here alongside with the canonical kinetic term. The
above, gives second order field equations due essentially to the
divergence free property of the Einstein tensor. The Einstein
scalar tensor term is one of the Fab 4 terms
\cite{Charmousis:2011bf}, having interesting implications on the
cosmological constant problem. Furthermore, on cosmological
backgrounds, this term with the function $G(\chi)$ constant, leads
to an accelerated expansion without the need of any scalar
potential, as noted for the first time in \cite{Amendola:1993uh}.
The presence of this coupling in the Lagrangian gives second-order
field equations  \cite{Amendola:2005cr} as part of a Kaluza Klein
reduction of Einstein-Gauss-Bonnet theory. These features
attracted much interest in inflationary cosmology
\cite{Sushkov:2009hk,germani}, particle production after inflation
\cite{Koutsoumbas:2013boa} and also late-time cosmology
\cite{acceleration}.

Local solutions for the action (\ref{EGBscalaraction})  and with
the coupling function $G$ constant were discussed in several
recent papers \cite{Rinaldi:2012vy}. Spherically symmetric black
hole solutions which are asymptotically anti-de Sitter were found.
They all rely on switching off the primary hair integration
constant by imposing a specific geometric Anzatz. This bifurcates
the no hair argument of \cite{Hui:2012qt} since it allows the
scalar field to be non-trivial. Unfortunately this is not  enough
to completely evade singular behaviour of the scalar field and
solutions are generically singular for the derivative of the
scalar field on the horizon. This is after all a common problem in
theories with a scalar field coupled to gravity \cite{BBMB}. To circumvent the
problem of regularity of local solutions one can
 break shift symmetry of the scalar field by introducing a mass term for
the scalar field \cite{Kolyvaris:2011fk,Kolyvaris:2013zfa}. Another way to remedy
this  problem, while keeping shift-symmetry,
\cite{Babichev:2013cya}, is to introduce an additional, mild,
linear dependence in the time coordinate for the scalar field (see
also the recent generalization of \cite{Kobayashi:2014eva}). This
yields an additional integration constant  while the shift
symmetry is essential in keeping the field equations
time-independent and consistent for a static space-time Anzatz.
This permits asymptotically flat (or de-Sitter) solutions and
crucially gives regular scalar tensor black holes
\cite{Babichev:2013cya}. Although the role of time dependence is
essential for regularity, the physical
significance associated to time dependence, as a genuine scalar
hair charge, is not as yet understood. An important question arises:  is it possible to generate an asymptotically flat
black hole with genuine primary scalar hair? This is one of the important
questions we will treat in this article.

Towards this end we will consider theories of the type
(\ref{EGBscalaraction}) where  the coupling function $G(\chi)$ will be non trivial.
The strategy we will follow will be to upgrade the coupling of the scalar-tensor
(\ref{EGBscalaraction}) interaction $G$, to full conformal invariance using results from Weyl geometry.
This will effectively introduce a second scalar field
and we will thus consider a bi-scalar-tensor theory.  We will then
 look for black hole solutions in close analogy to a scalar field conformally coupled to
 gravity \cite{BBMB}.

Gravity theories having  conformal invariance have many
advantages. Conformally invariant actions seem to play an
important role in early universe cosmology
\cite{Bars1}-\cite{Kallosh} and black hole physics \cite{thooft}.
Additionally it has been shown that the holographic
renormalisation procedure in four-dimensional General Relativity
(GR) can be achieved by adding a topological invariant term, which
in turn gives an on shell action equal to conformal gravity
\cite{Mis-Ole}. On the other hand it has also been proven that
starting from conformal gravity and requiring appropriate boundary
conditions an equivalence with Einstein GR can be made
\cite{Maldacena:2011mk}. Concerning local solutions   in scalar
tensor theories, adding a non-minimally coupled term of the scalar
field to gravity, respecting conformal invariance,  is the only
coupling allowing for black hole geometries \cite{BBMB}. However,
the resulting solutions are problematic since the scalar field
blows up at the event horizon. This pathology can be cured  by
adding  a cosmological constant alongside with a self-interaction
term of the scalar field
\cite{Ricardo,Anabalon:2012tu,Gonzalez:2013aca}. Another question we will successfully treat in this article will be to look for regular asymptotically flat
black holes of the BBMB type with primary hair.

Recently it has been proven \cite{Padilla:2013jza}, that in four
dimensional scalar tensor theories, the only combination which
respects conformal symmetry for a single scalar and has second
order field equations, comes in the form of the the well known
action \cite{deser} \be \int d^{4}x\sqrt{-g}\big[
-\frac{1}{2}\nabla_{\m}\phi\nabla^{\m}\phi-\frac{1}{12}\phi^{2}R-\a\phi^{4}\big]~.
\label{mtz} \ee  The full case for a single scalar theory in
arbitrary dimension is discussed in  \cite{Tsouk}.  Despite the
fact that the discussion so far has been in the context of a single
scalar case, we can also include higher numbers of scalars in our
theory. A wide class of actions that consider a multi-scalar
version of Horndeski's theory has been proposed in
\cite{Padilla:2012dx}. These theories do not possess full
generality \cite{Kobayashi:2013ina}, but in Minkowski space-time
they are the most general ones \cite{Sivanesan:2013tba}. Full
conformal invariance in these theories can be found by applying
the method of Ricci gauging \cite{Iorio:1996ad}.  In this way  an
efficient method has been presented, where at least for the
bi-scalar case we can have the most general Horndeski action
(\ref{horny}) promoted to an action having a full conformal
invariance  that avoids Ostrogradski ghosts
\begin{equation}\label{biscal}
S_{\text{local}}[\tilde{\phi}, \tilde{g}]=\int d^4 x
\sqrt{-\tilde{g}} \left[K(\tilde{\phi},
\tilde{X})-G_3(\tilde{\phi}, \tilde{X}) {\cal \tilde{E}}_1+G_4
(\tilde{\phi}, \tilde{X})\tilde{R}+G_{4, X} {\cal
\tilde{E}}_2+G_5(\tilde{\phi}, \tilde{X})
\tilde{G}_{\mu\nu}\tilde{\nabla}^\mu \tilde{\nabla}^\nu
\tilde{\phi}-\frac{G_{5, X}}{6} {\cal \tilde{E}}_3 \right]~,
 \end{equation}
where the composite fields $\tilde{\phi}=\phi/\pi$,
$\tilde{g}_{\mu\nu}=\pi^{2}g_{\mu\nu}$ have the following
transformation properties $g_{\m\n} \rightarrow \Omega(x)^2
g_{\m\n}, ~\pi\rightarrow \pi/\Omega(x), ~\phi \rightarrow
\phi/\Omega(x)$.

The work and our main results are organized as follows. In Section
\ref{secs.1} we will set up our bi-scalar tensor theory explaining
how the scalar field coupled to the Einstein tensor can be
transformed to a fully conformal invariant term with the addition
of an extra  scalar field $\phi$.  In Section \ref{secs.2} we will
introduce a cosmological constant and an electromagnetic field and
look for static black hole solutions. They will end up being of
planar horizon sections and will thus necessitate a negative
cosmological constant. In Section \ref{secs.3} we will allow for
the  scalar field $\phi$ to be coupled conformally to gravity and
find the according planar black hole. In Section \ref{secs.4}  in
order to render the solutions obtained in the previous sections
regular we will allow for the time dependence of the scalar field.
This will lead us to eventually consider a different coupling
function $G$ non conformally invariant but tailored to the lower
order BBMB action \cite{BBMB}. Then  for this theory we will
obtain two asymptotically flat black hole solutions of spherical
symmetry where the  BBMB scalar $\phi$ will be regular and have
genuine primary scalar hair.  Finally in Section \ref{secs.5} are
our conclusions and in the appendix we give some technical
details.

\section{setup}
\label{secs.1}

Before we  set up our theory we will review the black hole
solutions resulting from  a  scalar field conformally coupled to
gravity with an action given by (\ref{mtz}) supplemented by an
Einstein-Hilbert term with a cosmological constant.

The field equations are,
\bea G_{\mu \nu}+ \Lambda g_{\mu \nu}&=& 8 \pi T_{\mu \nu}~,
\label{ensequs}\\
\square \phi&=&\frac{1}{6}R\phi+4\alpha \phi^3~,\label{klenq} \eea
where the energy-momentum tensor is given by \be T_{\mu
\nu}=\partial_\mu \phi \partial_\nu \phi-\frac{1}{2}g_{\mu \nu}
g^{\alpha \beta}\partial_\alpha \phi \partial_\beta \phi
+\frac{1}{6}\Big{[}g_{\mu \nu}\square-\nabla_\mu \nabla_\nu
+G_{\mu \nu} \Big{]} \phi^2-g_{\mu \nu} \alpha \phi^4~. \ee and
$\square \equiv g^{\mu \nu}\nabla_\mu \nabla_\nu$.  Since
(\ref{mtz}) is invariant under conformal transformations \be
g_{\mu \nu}\rightarrow \Omega^2 (x) g_{\mu
\nu},\,\,\,\,\,\,\,\,\,\,\,\, \phi \rightarrow \Omega^{-1} (x)
\phi~,\label{tranf} \ee the stress tensor is traceless and as a
consequence, the scalar curvature is constant \be
R=4\Lambda~.\label{trace} \ee Adopting a particular spherically symmetric
ansatz for the metric \be
ds^{2}=-f(r)dt^{2}+\frac{1}{f(r)}dr^{2}+r^{2}d\Omega^{2}~, \ee
 because of the relation (\ref{trace}) we get, \be
f(r)=-\frac{\Lambda}{3}r^2+1+\frac{c_{1}}{r}+\frac{c_{2}}{r^{2}}~.\label{metfun}
\ee The  $1/r^{2}$ term appearing in (\ref{metfun}) is sourced
by the presence of the scalar field which plays a similar role to an EM field.

Solving the full system of equations (\ref{ensequs}) and
(\ref{klenq}) the constants $c_1$ and $c_2$ are specified and we
obtain the solution \be
ds^2=-\Big{[}-\frac{\Lambda}{3}r^2+\Big{(}1-\frac{GM}{r}\Big{)}^2\Big{]}dt^2
+\Big{[}-\frac{\Lambda}{3}r^2+\Big{(}1-\frac{GM}{r}\Big{)}^2\Big{]}^{-1}dr^2+r^{2}d\Omega^{2}~.
\ee this is known as the MTZ hairy black hole \cite{Ricardo}. It
is important to note that for a solution to exist a consistency
relation should hold connecting the constants of the theory
$\alpha=-\frac{2}{9}\pi \Lambda G$.  We note here that the
presence of the cosmological constant in the action from which the
field equations (\ref{ensequs}) and (\ref{klenq}) result makes the
scalar field regular on the horizon, hiding irregularities of the
scalar field behind the horizon. The black hole, not having an independent integration constant associated to the scalar field has secondary hair.

We will follow a similar strategy to find black hole solutions in
Horndeski theory (\ref{horny}) including in particular the term
\be \sqrt{-g}G(\chi)G^{\m\n}\nabla_{\m}\chi\nabla_{\n}\chi \ee
appearing in  (\ref{EGBscalaraction}). We will fix the coupling
function $G(\chi)$ and add additional terms in order to achieve
conformal invariance.  We will construct our conformally invariant
action in such a way that it  encodes various conformal weights
for one of the scalars \cite{Faci:2012yg}, through the use of Weyl
Geometry.

Let us suppose that the scalar field $\chi$ has conformal weight $w_{\chi}$ or
equivalently that it transforms as \be \chi \rightarrow
\Omega^{w_{\chi}}\chi~, \ee and the metric has the usual
transformation \be g_{\m\n} \rightarrow \Omega^{2}g_{\m\n}~. \ee
Now let us introduce a second  scalar field that transforms as \be
\phi \rightarrow \Omega^{-1} \phi~. \ee In order to construct a
conformally invariant action we need to promote the metric
$g_{\m\n}$ to $\tilde{g}_{\m\n}=\phi^{2}g_{\m\n}$ which is by
construction conformally invariant and modify the covariant
derivative that acts on $\chi$ as \be
D_{\m}\chi=\nabla_{\m}\chi-\frac{w_{\chi}}{2}\chi
\partial_{\m}\ln\phi^{-2}~. \ee Promoting the metric to the tilded
one   guarantees that they will be conformally invariant
\cite{Padilla:2013jza}, \cite{Tsouk} (see also \cite{Oliva:2011np}
for an alternative way to construct conformally coupled scalar
theories. This is done by introducing a four-rank tensor, that
transforms covariantly under local Weyl rescallings). Furthermore
the coupling function $G(\chi)$ has to take the form \be
G(\chi)=\chi^{-2}. \label{Gfunction}\ee It is easy to check that
$D_{\m}\chi \rightarrow \Omega^{w_{\chi}} D_{\m}\chi$, which
justifies the choice for the function $G(\chi)$ of
(\ref{Gfunction})\footnote{See \cite{Faci:2012yg} for a general
discussion on Weyl geometry and conformal invariance.}.

If we now want to make a conformally invariant action using the
Einstein tensor then one choice is the following \bea
\sqrt{-\tilde{g}}\,G(\chi) \tilde{G}^{\m\n}D_{\m}\chi D_{\n}\chi&=&\sqrt{g}\chi^{-2}\Big[G^{\m\n}\nabla_{\m}\chi\nabla_{\n}\chi+2w_{\chi}\frac{\chi}{\phi}G^{\m\n}\nabla_{\m}\chi\nabla_{\n}\phi+\left(w_{\chi}\frac{\chi}{\phi}\right)^{2}G^{\m\n}\nabla_{\m}\phi\nabla_{\n}\phi\nn \\
&&\qquad \qquad \quad +4\phi^{-2}\nabla^{\m}\phi\nabla^{\n}\phi\nabla_{\m}\chi\nabla_{\n}\chi+6 w_{\chi}\phi^{-3}\chi\nabla^{\k}\phi\nabla_{\k}\phi\nabla^{\m}\chi\nabla_{\m}\phi\nn \\
&&\qquad \qquad \quad +3\left(w_{\chi}\frac{\chi}{\phi}\right)^{2}\left(\nabla^{\m}\phi\nabla_{\m}\phi\right)^{2}-\phi^{-2}\nabla^{\k}\phi\nabla_{\k}\phi\nabla^{\m}\chi\nabla_{\m}\chi\nn \\
&&\qquad \qquad \quad -2\phi^{-1}\nabla^{\m}\nabla^{\n}\phi\nabla_{\m}\chi\nabla_{\n}\chi+2\phi^{-1}\square\phi\nabla^{\m}\chi\nabla_{\m}\chi\nn \\
&&\qquad \qquad \quad -4w_{\chi}\frac{\chi}{\phi^{2}}\nabla^{\m}\nabla^{\n}\phi\nabla_{\m}\chi\nabla_{\n}\phi+w_{\chi}\frac{\chi}{\phi^{2}}\square\phi\nabla^{\m}\chi\nabla_{\m}\phi\nn \\
&&\qquad \qquad \quad
-2\phi^{-1}\left(w_{\chi}\frac{\chi}{\phi}\right)^{2}\nabla^{\m}\nabla^{\n}\phi\nabla_{\m}\phi\nabla_{\n}
\phi+2\phi^{-1}\left(w_{\chi}\frac{\chi}{\phi}\right)^{2}\square\phi\nabla^{\m}\phi\nabla_{\m}\phi
\Big]~,  \nonumber \\ \eea which  by construction is conformally
invariant.{The $\tilde{G}^{\m\n},$ refers to the Einstein tensor constructed using the rescaled metric $\tilde{g}_{\m\n}$ and its inverse. A choice that simplifies the above expression is
 $w_{\chi}=0$. Setting $\Psi=\ln(\chi)$, we
have: \bea \label{zerowact}
\mathcal{L}=\sqrt{-\tilde{g}}G(\chi)\tilde{G}^{\m\n}D_{\m}\chi D_{\n}\chi&=&\sqrt{g}\Big[G^{\m\n}\nabla_{\m}\Psi\nabla_{\n}\Psi+4\phi^{-2}\nabla^{\m}\phi\nabla^{\n}\phi\nabla_{\m}\Psi\nabla_{\n}\Psi-\phi^{-2}\nabla^{\k}\phi\nabla_{\k}\phi\nabla^{\m}\Psi\nabla_{\m}\Psi \nn \\
&&\qquad \qquad \quad
-2\phi^{-1}\nabla^{\m}\nabla^{\n}\phi\nabla_{\m}\Psi\nabla_{\n}\Psi+2\phi^{-1}\square\phi\nabla^{\m}\Psi\nabla_{\m}\Psi\Big]
\eea and the action is shift symmetric in $\Psi$. This term which
is  conformally invariant will be our higher order ingredient for
our bi-scalar-tensor theory.

The field equations resulting from the   variation of
(\ref{zerowact}) with respect to $g_{\m\n},\Psi$ and $\phi$ are
given in the following.  \begin{itemize}

\item Field equation for the scalar field $\Psi$:

\bea
&-&2G^{\m\n}\nabla_{\m}\nabla_{\n}\Psi-8\nabla_{\m}\left( \phi^{-2}\nabla^{\m}\phi\nabla^{\n}\phi\nabla_{\n}\Psi\right)
+2\nabla_{\m}\left(\phi^{-2}\nabla^{\k}\phi\nabla_{\k}\phi\nabla^{\m}\Psi \right)
-4\phi^{-2}\nabla^{\m}\nabla^{\n}\phi\nabla_{\m}\Psi\nabla_{\n}\phi\nn\\
&&+4\phi^{-1}R_{\m\n}\nabla^{\m}\phi\nabla^{\n}\Psi+4
\phi^{-1}\nabla^{\m}\nabla^{\n}\phi\nabla_{\m}\nabla_{\n}
\Psi+4\phi^{-2}\square\phi\nabla^{\m}\phi\nabla_{\m}\Psi-4\phi^{-1}\square\phi\square\Psi=\mathcal{E}_{\Psi}~.
\label{cijpsi} \eea

\item Field equation for the scalar field $\phi$:

\bea
&-&8\phi^{-3}\nabla^{\m}\phi\nabla^{\n}\phi\nabla_{\m}\Psi\nabla_{\n}\Psi-8\nabla^{\m}\left( \phi^{-2}\nabla^{\n}\phi\nabla_{\m}\Psi\nabla_{\n}\Psi\right)+2\phi^{-3}\nabla^{\k}\phi\nabla_{\k}\phi\nabla^{\m}\Psi\nabla_{\m}\Psi\nn\\
&&+2\nabla_{\k}\left(\phi^{-2}\nabla^{\k}\phi\nabla^{\m}\Psi\nabla_{\m}\Psi \right)+4\phi^{-2}\nabla^{\m}\nabla^{\n}\phi\nabla_{\m}\Psi\nabla_{\n}\Psi-4\phi^{-2}\square\phi\nabla^{\m}\Psi\nabla_{\m}\Psi\nn\\
&&-4\phi^{-3}\nabla^{\m}\phi\nabla^{\n}\phi\nabla_{\m}\Psi\nabla_{\n}\Psi+4\phi^{-3}\nabla^{\n}\phi\nabla_{\n}\phi\nabla^{\m}\Psi\nabla_{\m}\Psi-4\phi^{-2}\nabla^{\m}\phi\nabla^{\n}\Psi\nabla_{\m}\nabla_{\n}\Psi\nn\\
&&+4\phi^{-2}\square\Psi\nabla^{\m}\phi\nabla_{\m}\Psi-2
\phi^{-1}(\square\Psi)^{2}
+2\phi^{-1}\nabla_{\m}\nabla_{\n}\Psi\nabla^{\m}\nabla^{\n}\Psi+2\phi^{-1}R^{\m\n}\nabla_{\m}\Psi\nabla_{\n}\Psi=\mathcal{E}_{\phi}
\label{cijphi}~. \eea

\item Metric field equations :
\bea
&-&\frac{1}{2}g_{\m\n}G^{\a\b}\nabla_{\a}\Psi\nabla_{\b}\Psi+2G_{(\m}^{\,\,\,\,\,\la}\nabla_{\n)}\Psi\nabla_{\la}\Psi+\frac{1}{2}R\nabla_{\m}\Psi\nabla_{\n}\Psi-\frac{1}{2}R_{\m\n}\nabla^{\a}\Psi\nabla_{\a}\Psi\nn\\
&&+\frac{1}{2}g_{\m\n}\left((\square\Psi)^{2}-\nabla_{\a}\nabla_{\b}\Psi\nabla^{\a}\nabla^{\b}\Psi-R_{\a\b}\nabla^{\a}\Psi\nabla^{\b}\Psi\right)+\nabla_{\m}\nabla^{\a}\Psi\nabla_{\n}\nabla_{\a}\Psi-\square\Psi\nabla_{\m}\nabla_{\n}\Psi\nn\\
&&+R_{\m\,\,\,\n}^{\,\,\,\a\,\,\,\,\b}\nabla_{\a}\Psi\nabla_{\b}\Psi-2g_{\m\n}\phi^{-2}\nabla^{\a}\phi\nabla^{\b}\phi\nabla_{\a}\Psi\nabla_{\b}\Psi+8\phi^{-2}\nabla^{\k}\phi\nabla_{\k}\Psi\nabla_{(\m}\phi\nabla_{\n)}\Psi\nn\\
&&+\frac{1}{2}g_{\m\n}\phi^{-2}\nabla^{\k}\phi\nabla_{\k}\phi\nabla^{\la}\Psi\nabla_{\la}\Psi-\phi^{-2}\nabla_{\m}\phi\nabla_{\n}\phi\nabla^{\k}\Psi\nabla_{\k}\Psi-\phi^{-2}\nabla^{\k}\phi\nabla_{\k}\phi\nabla_{\m}\Psi\nabla_{\n}\Psi\nn\\
&&+g_{\m\n}\phi^{-1}\nabla^{\a}\nabla^{\b}\phi\nabla_{\a}\Psi\nabla_{\b}\Psi-4\phi^{-1}\nabla_{(\m}\nabla^{\k}\phi\nabla_{\n)}\Psi\nabla_{\k}\Psi+2\nabla_{\k}\left(\phi^{-1}\nabla_{(\m}\phi\nabla_{\n)}\Psi\nabla^{\k}\Psi\right)\nn\\
&&-\nabla_{\a}\left(\phi^{-1}\nabla^{\a}\phi\nabla_{\m}\Psi\nabla_{\n}\Psi\right)-g_{\m\n}\phi^{-1}\square\phi\nabla^{\a}\Psi\nabla_{\a}\Psi+2\phi^{-1}\nabla_{\m}\nabla_{\n}\phi\nabla^{\a}\Psi\nabla_{\a}\Psi\nn\\
&&+2\phi^{-1}\square\phi\nabla_{\m}\Psi\nabla_{\n}\Psi-2\nabla_{(\m}\left(\phi^{-1}\nabla_{\n)}\phi\nabla^{\b}\Psi\nabla_{\b}\Psi\right)
+g_{\m\n}\nabla_{\a}\left(\phi^{-1}\nabla^{\a}\phi\nabla^{\b}\Psi\nabla_{\b}\Psi\right)=\mathcal{E}_{\m\n}~.
\label{cijgmn} \eea

\end{itemize}

As can be seen from the above expressions we have labeled the variation of (\ref{zerowact}) with respect the scalar fields $\Psi$,$\phi$ and the metric $g_{\m\n}$, as $\mathcal{E}_{\Psi},\mathcal{E}_{\phi}$ and $\mathcal{E}_{\m\n}$ respectively. It can be seen that the trace of (\ref{cijgmn}) is identically
zero by virtue of the  scalar field equation (\ref{cijphi}).
Despite that these expressions are long and complicated, we will
see that they can be tractable. In the following sections we will
solve the equations (\ref{cijpsi})-(\ref{cijgmn}) for various
cases.

\section{Planar black holes with a cosmological constant and a Maxwell Field}
\label{secs.2}

Consider the following action \be S=\int d^{4}x
\sqrt{-g}\,\frac{R-2\,\Lambda}{16\p G_{N}}-\gamma \int d^{4}x
\sqrt{-g}\,\frac{1}{16\pi}F^{\m\n}F_{\m\n}+\b\,\int d^{4}x
\sqrt{-\tilde{g}}\,\tilde{G}^{\m\n}D_{\m}\Psi D_{\n}\Psi~,
\label{action4} \ee where $\gamma$ is a dimensionless coupling
constant used to switch on and off the Maxwell field, and the last term is by construction conformally invariant (section \ref{secs.1}).
The field equations for $\Psi$ and
$\phi$ are respectively (\ref{cijpsi}) and (\ref{cijphi}), while
for the metric we have \be\label{fullgeqs2}
H_{\m\n}=\frac{1}{16\p\,G_{N}}\big(G_{\m\n}+\Lambda\,g_{\m\n}\big)+\gamma\big(\frac{1}{32\pi}g_{\m\n}
F^{\a\b}F_{\a\b}-\frac{1}{8\pi}F_{\m}^{\,\,\,\a}F_{\n\a}\big)+\b\mathcal{E}_{\m\n}=0~.
\ee The metric ansatz we consider is the following \be
ds^{2}=-f(r)dt^{2}+\frac{1}{f(r)}dr^{2}+r^{2}d\Omega^{2}_{\kappa}~,
\ee where
\be\label{basemet}
d\Omega^{2}_{\kappa}=\frac{1}{1-\k
x^{2}}dx^{2}+x^{2}dy^{2}~.
 \ee
  Taking the trace of
(\ref{fullgeqs2}) and using the field equation for $\phi$, we get
$R=4\La$ from which we get,
\be
f(r)=\k+\frac{c_{1}}{r}+\frac{c_{2}}{r^{2}}-\frac{\Lambda}{3}\,r^{2}~.\label{metcons}
\ee
Although conformal symmetry nicely (seems to) close in to a solution, the above is not in general a solution for the theory (\ref{action4}).
\subsection{Planar stealth solution}

With the inclusion of the cosmological constant (we set
$\gamma=0$), we consider the metric function with a flat
horizon $\kappa=0$ \be f(r)=\frac{r^2}{\lambda^2}-\frac{m}{r} \
, \ee where $\Lambda=-3/\lambda^2$ Then from $H_{r}^{\,\,\,r}=0$ we get
\be
H_{r}^{\,\,\,r}=0=>\big(\phi+r\,\phi'\big)\,\big(3m\,\phi'+r^{2}\Lambda\,\big(\phi+r
\phi'\big)\big)=0~.
\ee
If the first bracket is zero, then the
field equations are satisfied with
 \bea
\phi(r) & = & \frac{c_0}{r} \ ,\label{solut1}\\
\Psi(r) & = & \tilde{C}_2 + \tilde{C}_1
\ln\left(r^{3/2}+\sqrt{r^3-m^2 \lambda ^2}\right)~.\label{solut2}
\eea
Notice that the scalar $\phi$ in (\ref{solut1}) is
regular everywhere apart from the origin and we have a scalar charge $c_0$.
The scalar  $\Psi$ is regular up to and including the horizon at $r=r_h$, where $f(r_h)=0$. Its derivative however is
divergent and the scalar $\Psi$ is imaginary for $r<r_h$.
Calculating the on shell
action, particularly for the
$\sqrt{-\tilde{g}}\,\tilde{G}^{\m\n}D_{\m}\Psi D_{\n}\Psi$ term,
we can see that it is actually zero and therefore regular on shell. This irregularity of the scalar field encountered in similar solutions \cite{Rinaldi:2012vy} can be completely remedied by including time dependence in the manner of \cite{Babichev:2013cya} as we will see in a forthcoming section.

Now if the second bracket is zero then the field equations are satisfied with
\bea
\phi(r) & = & \frac{C_0}{(-3m -r^{3}\,\Lambda)^{1/3}}~,\label{solut3}\\
\Psi(r) & = & C_2 + \frac{2\,C_1}{3\sqrt{\Lambda}}
\ln\Big[2\left(r^{3/2}\Lambda+\sqrt{\Lambda}\sqrt{r^3\Lambda+3m}\right)\Big]~. \eea
Here however, the solution (\ref{solut3}) for the scalar field $\phi$
blows up at the horizon. If we want to have regular behaviour of
the scalar field on the horizon then the scalar charge $C_0$ has to be
zero. In this case the action (\ref{action4}) (with $\gamma=0$)
has a local solution with one regular scalar field. Note that in
this case the derivative coupling of the scalar field to the
Einsten tensor is not any more conformally invariant.

\subsection{Electric solution with a flat horizon}

We now switch back on the Maxwell field ($\g=1$).  The field equations
for the metric are \be\label{fullgeqs3}
H_{\m\n}=\frac{1}{16\p\,G_{N}}\big(G_{\m\n}+\Lambda\,g_{\m\n}\big)+\frac{1}{32\pi}g_{\m\n}F^{\a\b}
F_{\a\b}-\frac{1}{8\pi}F_{\m}^{\,\,\,\a}F_{\n\a}+\b\mathcal{E}_{\m\n}=0~.
\ee From the Maxwell equation we obtain \bea
\nabla_{\m}\,F^{\m\n}&=&0~,\\
A(r)&=&-\frac{Q}{r}~. \eea From the trace of the field equations
for $\k=0$ we find that the metric function is \be
\label{elplanar}
f(r)=\frac{c_{1}}{r}+\frac{c_2}{r^{2}}-\frac{\Lambda}{3}\,r^{2}.
\ee In this case the equations are satisfied only if $c_{2}=0$, or
if $c_{2}=Q^{2}\,G_{N}$. If we choose $c_{2}=0$ then the electric charge
$Q$ must also be zero, so we end up to the previous solution that
we have already discussed. So $c_{2}=Q^{2}\,G_{N}$ is the only
choice. Setting this value for $c_{2}$, from the $H_{r}^{\,\,r}=0$
equation we get \be
H_{r}^{\,\,\,r}=0=>\big(\phi+r\,\phi'\big)\,\big(-3r^{2}c_{1}\,\phi'+Q^{2}\,G_{N}(\phi-3r\phi')+r^{4}\Lambda\,\big(\phi+r
\phi'\big)\big)=0~. \label{choice}\ee Choosing the first bracket to be zero we
obtain \be \phi(r)=\frac{c_{0}}{r}~. \ee Substituting the value of
$\phi$ back to the field equations we can solve for $\Psi'$
 \be
\Psi'(r)=\frac{r^{3/2}C_{1}}{\sqrt{3rc_{1}+4Q^{2}G_{N}}\,\sqrt{-3rc_{1}-3Q^{2}G_{N}+r^{4}\Lambda}}~.
\ee Because of the shift symmetry  we only need $\Psi'$ and $\Psi''$ since only these
expressions appear in the field equations.  If we calculate the
value of $\sqrt{-\tilde{g}}\,\tilde{G}^{\m\n}D_{\m}\Psi
D_{\n}\Psi$ we get once again zero as a result and therefore the fact that $\Psi'$ is divergent on the horizon is a somewhat milder iregularity.

Now if the second bracket is zero we find that the solution for
$\phi$ is \be
\phi(r)=\frac{c_{0}\,r^{1/3}}{(-3rc_{1}-3Q^{2}G_{N}+r^{4}\Lambda)^{1/3}}~,
\ee and again \be
\Psi'(r)=\frac{r^{3/2}C_{1}}{\sqrt{3rc_{1}+4Q^{2}G_{N}}\,\sqrt{-3rc_{1}-3q^{2}G_{N}+r^{4}\Lambda}}~.
\ee
The introduction of a Maxwell field fixes the value of $c_{2}$ in (\ref{metcons}) to
be proportional to the charge. However, as it happens in the case
of a constant coupling constant \cite{Rinaldi:2012vy} the scalar
fields are not regular on the horizon. It seems that another scale
is needed in order to hide irregularities behind the black hole
horizon. The on-shell value of $\sqrt{-\tilde{g}}\,\tilde{G}^{\m\n}D_{\m}\Psi
D_{\n}\Psi$ is again vanishing.

\section{Introducing a conformally coupled scalar $\phi$}
\label{secs.3}

In the previous section we saw that the scalar field $\phi$ is either regular or irregular on the horizon depending on the branch we choose (\ref{choice}). Since we want to eventually close in on BBMB type solutions \cite{BBMB} we will now assume that the scalar field $\phi$ has a conformal coupling to gravity. We consider therefore
following action \be S=\int d^{4}x
\sqrt{-g}\,\Big[\frac{R-2\,\Lambda}{16\p
G_{N}}-\frac{1}{2}g^{\m\n}\nabla_{\m}\phi\,\nabla_{\n}\phi-\frac{1}{12}\,\phi^{2}\,R-\a\,\phi^{4}-\,\frac{1}{16\pi}F^{\m\n}F_{\m\n}\Big]+\b\,\int
d^{4}x \sqrt{-\tilde{g}}\,\tilde{G}^{\m\n}D_{\m}\Psi D_{\n}\Psi~.
\ee Then the field equations for the metric become
\be\label{fullgeqs4}
H_{\m\n}=\frac{1}{16\p\,G_{N}}\big(G_{\m\n}+\Lambda\,g_{\m\n}\big)-T_{\m\n}^{(\phi)}
+\frac{1}{32\pi}g_{\m\n}F^{\a\b}F_{\a\b}-\frac{1}{8\pi}F_{\m}^{\,\,\,\a}F_{\n\a}+\b\mathcal{E}_{\m\n}=0~,
\ee where \be
T_{\m\n}^{(\phi)}=\frac{1}{2}\nabla_{\m}\phi\nabla_{\n}\phi-\frac{1}{4}g_{\m\n}\,\nabla^{\a}\phi\,\nabla_{\a}\phi
+\frac{1}{12}\big[g_{\m\n}\square-\nabla_{\m}\nabla_{\n}+G_{\m\n}\big]\phi^{2}-\frac{1}{2}g_{\m\n}\,\a\,\phi^{4}~,
\ee and the field equation for the scalar field $\phi$ changes
accordingly from (\ref{cijphi}) to \be
\square\,\phi-\frac{1}{6}\,\phi\,R-4\,\a\,\phi^{3}+\b\,\mathcal{E}_{\phi}=0~.
\ee Setting again $\k=0$ we find the metric function \be
f(r)=\frac{c_{1}}{r}+\frac{c_{2}}{r^{2}}-\frac{\Lambda}{3}\,r^{2}~,
\ee while the Maxwell field reads,
 \be
A(r)&=&-\frac{Q}{r}~. \ee
Now if we choose the scalar field
to be $\phi(r)=\frac{c_{0}}{r}$, then the equation
$H_{t}^{\,\,t}-H_{r}^{\,\,r}=0$ is automatically satisfied. We can
now take the $H_{t}^{\,\,t}-H_{x}^{\,\,x}=0$, ($x$ is a coordinate
in the base manifold whose curvature $\k$ we have set to zero) and
solve for $\Psi'$ \bea
&&H_{t}^{\,\,t}-H_{x}^{\,\,x}=0=>\nn \\
&&-3r^{2}c_{2}+3\big(Q^{2}r^{2}+2\pi\,c_{0}^{2} (r\,c_{1}+2c_{2}) \big) G_{N}-2\pi\,\b G_{N}\big(9r^{2}c_{1}^{2}+4c_{2}(9c_{2}+r^{4}\Lambda)+6c_{1}(7 r\,c_{2}+r^{2}\Lambda) \big)\Psi'^{2}\nn\\
&&-4\pi r\,\b (3 r c_{1}+4c_{2})\,G_{N}\,\big(-3(r
c_{1}+c_{2})+r^{4}\Lambda \big)\Psi'\,\Psi''=0=> \eea \be
\Psi'(r)=\pm\sqrt{\frac{3r^{2}c_{2}-\big(r^{2}(3Q^{2}-2\pi
r\,\b\,C_{1})+\pi\,c_{0}^{2}(3r\,c_{1}+4c_{2})
\big)G_{N}}{2\pi\,\b\,(3r\,c_{1}+4c_{2}) G_{N}
\big(-3(r\,c_{1}+c_{2})+r^{4}\Lambda \big)}}~. \ee Plugging back
the solution for $\Psi$ (the sign of $\Psi$ is irrelevant
according to the field equations) we have the following constraint
for the integration constant $c_{2}$ \be
c_{2}=(Q^{2}+8\pi\,\a\,c_{0}^{4})\,G_{N}~. \ee This is a
consistency relation between the constants and it is analogue to
the consistency relations that  appear in the usual conformally
coupled scalar field solutions, \cite{Ricardo}. Similar to what we
have seen in  the previous sections, the value of
$\sqrt{-\tilde{g}}\,\tilde{G}^{\m\n}D_{\m}\Psi D_{\n}\Psi$ is
vanishing. This solution is analogous to the planar black hole
found in \cite{Bardoux:2012tr}.  There it was found that in order
to support a planar MTZ black hole \cite{Ricardo}, one needed to
include two axionic fields. Here the same role is played by the
Galileon $\Psi$ although note that the space-time metrics are
quite different.

The above solution may not be the most general, since there was no
systematic way to find $\phi$. We started from an obvious ansatz
for $\phi$ which at the end turns out to be a correct one.

Additionally if we start from
\be
\phi(r)=\frac{c_{0}\,r^{1/3}}{(-3rc_{1}-3Q^{2}G_{N}+r^{4}\Lambda)^{1/3}}
\ee
and we set $\a=0$ and $c_{2}=Q^{2} G_{N}$ we see that
\be
\Psi(r)=\pm\frac{r^{3/2\sqrt{C1-\frac{2\b c_{0}^{2}(3rc_{1}+4Q^{2}G_{N})}{r^{1/3}(-3rc_{1}-3Q^{2}
G_{N}+r^{4}\Lambda)^{2/3}}}}}{2\b\sqrt{3rc_{1}+4Q^{2}G_{N}}\sqrt{-3rc_{1}-3Q^{2}G_{N}+r^{4}\Lambda}}
\ee
solves the equations.

 Again we observe irregularity of the scalar fields on the black
hole horizon, while also the value of $\tilde{G}^{\m\n}D_{\m}\Psi
D_{\n}\Psi$ is once again zero.

\section{Introducing linear dependence and regularity}
\label{secs.4}

In this section in order to address  the irregularity of the
galileon field on the horizon we will introduce a time dependent scalar field in the manner described in \cite{Babichev:2013cya}. As
before, we will first study the case of a $\Psi$-dependent
derivative coupling (\ref{zerowact}). We will then, using the construction ideas of  \cite{Babichev:2013cya},
extend the action to include a particular form of energy momentum tensor. In this way we will obtain asymptotically flat and regular solutions.

\subsection{Regular planar black hole}

Start as before with  (\ref{action4}) without an electromagnetic
field ($\gamma=0$). For the  metric we consider the general planar
ansatz
 $\kappa=0$, \be
ds^{2}=-h(r)dt^{2}+\frac{1}{f(r)}dr^{2}+r^{2}d \Omega  ^{2}~, \ee
where \be d\Omega ^{2}=dx^{2}+x^{2}dy^{2}~. \ee Consider now that
the scalar field has also a linear dependence in time,
$\Psi(t,r)=q t+ \psi(r)$  \cite{Babichev:2013cya}, \cite{Babichev:2010kj}. The field equations, due to the shift
symmetry of the field $\Psi$ will still be ODE's. Furthermore, the
field equation $H_{tr}=0$ is now non trivial and controls the flux
of the scalar field $\Psi$ which is time dependent while the
metric is forced to be static. Verifying this equation actually
also kills the scalar field equation for the $\Psi$ field. From the latter equation,
apart from the obvious solution $\Psi=const$, we get two other
possible solutions for $\phi$ \be \label{pos1} \phi=\frac{m}{r}~,
\ee as before (and regular away from $r=0$) and \be \label{pos2}
\phi=\frac{m}{(r h(r))^{1/3}}. \ee We reject the latter for it
will be singular whenever $h(r)=0$. We can then immediately solve
for $H_{tt}=0$ to obtain, \be
f(r)=\frac{r^2}{\lambda^2}-\frac{m}{r} ~, \ee and then inputting
the result in the $H_{rr}$ component we obtain $h(r)=f(r)$.
Finally the $H_{\theta\theta}=0$ equation gives us $\Psi(t,r)$ to
be \be \Psi(t,r)=q t\pm \int
\frac{\sqrt{q^2+\frac{C}{l^2}h(r)}}{h(r)}dr~. \ee
 The solution we found in Section~\ref{secs.2} is simply obtained for $q=0$.
 Going to Eddington-Finkelstein (EF) coordinates
\begin{equation}\label{v1}
    v = t + \int \frac{ dr}{f(r)}~,
\end{equation}
one finds as usual a regular future chart
\begin{equation}\label{metricEF}
    ds^2 = - f(r) dv^2 +2 \, dv dr + r^2 d\Omega^2~.
\end{equation}
Applying the same transformation for $\psi$ gives
\be\label{psiEF1} \psi(v,r)=q v+\frac{C}{l^2} \int
\frac{dr}{q\pm\sqrt{q^2+\frac{C}{l^2}h(r)}} \ee which is regular
at the future horizon for $\psi$ only for the plus branch of
(5.6). The minus sign in the denominator of (\ref{psiEF1}) is
excluded, since it can cause infinities. Hence by including a time
dependent scalar field $\psi$ we obtain a regular planar black
hole solution. Note that the scalar $\psi$ still has a light like
singularity for $v\rightarrow \infty$ but that is independent of
the black hole solution and the derivative of $\psi$ appearing in
the action is constant.

It is straightforward to switch on  the electric charge in the
action and obtain the electric  version of the solution with
$\psi(t,r)=qt+\psi(r)$. The metric has the previous form
(\ref{elplanar}) with $\phi=c_0/r$ but the scalar field $\psi(r)$
is quite more involved. The regularity mechanism works in the same
way.

\subsection{Constructing an asymptotically flat hairy black hole}

Up to now we have obtained solutions with a negative cosmological
constant and a planar horizon. Our action (\ref{action4}) was
constructed so as to have conformal symmetry in the higher order term
(\ref{zerowact}). We would obviously like to extend our results to
get solutions in asymptotically flat space-time. In order to do we will use the insight gained in the previous sections
and consider a slightly different action tailoring it to the construction method of
\cite{Babichev:2013cya}.
Consider therefore the following action, \be \label{extended
action} S=S_0+S_1 \ee where \be \label{bek} S_0=\int dx^{4}
\sqrt{-g}\; \left[\z R+\eta
\left(-\frac{1}{2}(\partial\phi)^{2}-\frac{1}{12}\phi^{2}R\right)\right]
\ee and \be S_1=\int dx^{4} \sqrt{-g}\; \left(\b
G_{\m\n}\nabla^{\m}\Psi\nabla^{\n}\Psi-\g
T_{\m\n}\nabla^{\m}\Psi\nabla^{\n}\Psi\right) ~, \label{johny} \ee
where $T_{\m\n}(\phi, g_{\mu\nu})$ is precisely the energy
momentum tensor of a scalar field $\phi$ conformally coupled to
gravity, \be
T_{\m\n}=\frac{1}{2}\nabla_{\m}\phi\nabla_{\n}\phi-\frac{1}{4}g_{\m\n}\nabla_{\a}\phi\nabla^{\a}\phi
+\frac{1}{12}\left(g_{\m\n}\square-\nabla_{\m}\nabla_{\n}+G_{\m\n}
\right)\phi^{2}~. \ee Here, only the part of the action $S_0$
multiplied with $\eta$ is conformally invariant. Moreover contrary
to the previous cases we do not need to set any conformal weight
for $\Psi$, which means that not only $\b$ is dimension-full, but
also $\g$. The reason for considering such an action is the
following. Consider the $\Psi$ field equation (obtained from
(\ref{johny})) which due to shift symmetry can be nicely written
as a current conservation equation, \be
\label{psii}
\nabla_{\m}J^{\m}=0~,
\qquad J^{\m}=\left(\b G^{\m\n}-\g T^{\m\n}
\right)\nabla_{\n}\Psi~. \ee Then note that the current vector
$J^\mu$ "contains" the metric field equations of the BBMB action
(\ref{bek}). As such we can refer to action $S_0$ as being
precursor of the higher order action $S_1$. We will see that this
will enable to obtain our desired result even though the field
equations associated to (\ref{extended action}) are very complex.
We give the field equations for the scalar field $\phi$ and the
metric, in the appendix and let's denote them ${\mathcal
E}_{\phi}=0$ and ${\mathcal H}_{\m\n}=0$ respectively.

We now proceed to adopt a spherically symmetric anzatz \be
ds^{2}=-h(r)dt^{2}+\frac{dr^{2}}{f(r)}+r^{2}d\Omega^{2}~, \ee
where $d\Omega^{2}$ is the line element for the 2-sphere, while
for the scalar fields we set \be \phi=\phi(r)\,\,\,
\text{and}\,\,\, \Psi=\Psi(t,r)=q t+\psi(r)~. \ee
Assuming that 
\be \label{teos}
 \b G_{rr}-\g T_{rr}=0~,
  \ee kills
the dangerous $J^r$ component of the current satisfying the
regularity requirement of \cite{Hui:2012qt} without imposing a
trivial $\psi$-field \cite{Babichev:2013cya},
\cite{Sotiriou:2013qea}. Equation (\ref{teos}), is a solution to $\mathcal H_{tr}=0$. In
fact the primary  hair charge{\footnote{This is the integration
constant emanating from the scalar equation (\ref{psii})}}
associated to the Galileon $\Psi$ is in this way set to zero. Now
having an extra scalar, $\phi$ we may hope to keep the primary
hair charge associated to that field. Indeed this is what happens;
we solve (\ref{teos}) for the function $f(r)$,
 \be\label{fheq}
 f(r)=\frac{\phi^2(r) h(r) (12\b-\g\,\phi^{2}(r))}{12\b \phi^2(r) (r\,h(r))'-\g\,(r\,\phi(r))'\,(r\,\phi^3(r)\, h(r))'}~.
\ee The above result for $f(r)$ satisfies the field equation of
$\Psi$ and the flux equation ${\mathcal H}_{tr}=0$.  Then we turn
to ${\mathcal H}_{rr}=0$ and solve for $\psi'(r)$, \be\label{psi1}
\psi'=\pm\frac{\sqrt{q^{2}
(-12\b+\g\,\phi^{2})h'(-12\b\,r+\frac{\g}{2}
(\phi^{2}r^{2})')-12(\g\z-\b\eta)(\frac{(\phi^{2})'}{4}(h^{2}r^{2})'+3h^{2}r\,\phi'
(r\phi)'})}{h (-12\b+\g\,\phi^{2})}~. \ee One can attempt a brute
force resolution of the final equation ${\mathcal H}_{tt}=0$ in
order to obtain the metric component $h(r)$. We did not manage to
solve this equation in all generality. Guided however by our previous form of the scalar field
$\phi$, namely  (\ref{pos1}) and (\ref{pos2}) we note that they
simplify considerably (\ref{fheq}). We choose therefore to impose
the former $\phi=c_0/r$ for it is regular apart from $r=0$ whereas
the latter is singular for $h=0$, the location of a possible event
horizon. The integration constant $c_0$ is the scalar hair associated to $\phi$.
We get that \be f(r)=\frac{h(r)\,(12 \b r^2-\g c_{0}^{2})}{12\b
r^2(\,r\,h(r))'}~. \ee and \be \psi'(r)=\mp q\frac{\sqrt{6 r
(\g\z-\b\eta)c_{0}^{2}(h^2(r)\,
r^2)'+(12\,r^{2}\,\b-\g\,c_{0}^{2})h'(r) 12\b
r^3}}{h(r)(12\,r^{2}\,\b-\g\,c_{0}^{2})}~. \ee Going back to the
equation ${\mathcal H}_{t}\,\,^{t}=0$ and substituting all of the
above expressions we see that setting, \be
h(r)=-\frac{\m}{r}+\frac{1}{r}\int
\frac{k(r)}{12\,r^{2}\,\b-\g\,c_{0}^{2}}dr~, \ee then equation
${\mathcal H}_{t}\,\,^{t}=0$ reduces to just an algebraic equation
for $k(r)$, \be \label{tsoukalovits}
q^{2}\b(-12r^{2}\,\b+\g\,c_{0}^{2})^{2}+k(r)\left(-24r^{2}\b\z+(\g\z+\b
\eta)c_{0}^{2}\right)-C_{1}\,r\,k^{3/2}=0~, \ee where $C_{1}$ is
an integration constant. Any solution to the above cubic gives a
solution to the full system of equation, retracing each field step
by step. Generically, these solutions will not have usual
asymptotics and will be quite tedious to write down.

Let us therefore look at some particular solutions. Start by assuming
$f=h$. This immediately leads us, using (\ref{tsoukalovits}) to,
\bea
&&f(r)=h(r)=1-\frac{m}{r}+\frac{\g c_{0}^{2}}{12\b r^{2}}~,\\
&&\phi(r)=\frac{c_{0}}{r}~,\\
&&\psi'(r)=\pm q \frac{\sqrt{mr-\frac{\g c_{0}^{2}}{12\b}}}{r \, h(r)},\\
&& \b \e+\g (q^2\b-\z)=0~. \eea and $C_1^{2}=12 \beta(\b
q^2-2\z)^{2}$. We see that we have a black hole solution with
primary scalar charge $c_0$ playing a similar role to an EM gauge
field for the metric solution. Going as before to EF coordinates
(\ref{v1}) we see that, \be\label{psiEF2} \psi=q v- q \int
\frac{dr} { 1\pm\sqrt{1-h(r)}} \ee which is regular at the outer
event horizon $h(r_h)=0$ (once again only the plus branch of
(\ref{psiEF2}) is taken). We have therefore obtained a scalar
tensor black hole which has primary hair charge $c_0$ for the
scalar field $\phi$ while it has a regular scalar Galileon field
$\psi$ with an additional charge $q$. The geometry of the black
hole is similar to that of Reissner-Nordstrom.

The second possibility we consider is to take $C_1=0$ in
(\ref{tsoukalovits}). For simplicity we impose $\b \e-\g \z=0$ and
we get, \be
h(r)=1-\frac{m}{r}, \qquad f(r)=(1-\frac{m}{r})\left(1-\frac{\g
c_0^2}{12 \b r^2} \right) \ee and \be\label{psi1} \psi'= \mp
\frac{q}{h(r)}
\sqrt{\frac{m\,r}{r^2-\frac{\g\,c_0^{2}}{12\beta}}}~. \ee Taking
$\gamma \beta<0$ is enough-for example-to ensure that the only relevant zero
of $h(r)$ and $f(r)$ is $r_h=m$. Otherwise we need $m>\sqrt{\frac{\g c_0^2}{12 \b}}$. Then
again, as before the EF chart, which in this case reads as
\begin{equation}\label{v}
    v = t + \int \frac{ dr}{\sqrt{f(r)h(r)}}~,
\end{equation}
is  regular and we get, \be \psi=q v- q \int
\frac{dr}{\sqrt{\left(1-\frac{\g c_0^2}{12 \b r^2}
\right)}(1\mp\sqrt{\frac{m}{r}})}~. \ee Again the plus branch is
the acceptable one.

\section{Conclusions}
\label{secs.5}

In this paper we have studied certain bi-scalar-tensor theories and found hairy black hole solutions. The theories we have investigated are characterized by the
 property that part of the action is conformally invariant.
As a result one scalar is a Galileon field with shift symmetry, while the latter is coupled conformally to part of the action. Our aim has been to combine older techniques developed for conformally coupled scalars \cite{BBMB}, \cite{Ricardo} with newer ones \cite{Babichev:2013cya} with the ultimate goal of finding black holes with primary hair.  This was eventually achieved, in a step by step manner in the last section.

The bi-scalar tensor theories we
studied emerge from the general scalar-tensor  theory put forward
by Hordenski \cite{horny}. Indeed taking one of the Horndeski scalar-tensor
interaction terms $G(\Psi) G^{\mu\nu}\nabla_\mu\Psi
\nabla_\nu\Psi$ and requiring conformal invariance with the help
of an extra scalar field $\phi$ we constructed conformally
invariant terms which yield second order field equations. Fixing the coupling $G(\Psi)$  in this way, our aim was to construct a theory where regular
solutions may be found due to the underlying conformal symmetry
much like in the case of the BBMB and MTZ solutions \cite{BBMB},
\cite{Ricardo}.  In order to do so, we had the freedom to  choose
the conformal weight of our fields without the need of adding
canonical  kinetic terms. These theories are a subset of the full
general conformally invariant bi-scalar theories  that have
 been recently introduced in \cite{Padilla:2013jza}.

Having fixed the higher order conformal term (\ref{zerowact}) with
a particular simplifying conformal weight we added to the action
an Einstein-Hilbert, a cosmological constant and an
electromagnetic field breaking the conformal invariance for
the full theory.  Indeed, had we worked with a theory admitting conformal invariance the solutions would have had one free function as a result of the symmetry.
We also considered the case of an additional
conformal coupling of the second scalar $\phi$ in the manner of
\cite{BBMB}. In all cases we found anti de Sitter planar black
holes where the second scalar, akin in some cases to MTZ
\cite{Ricardo}, was everywhere regular apart from the black hole
singularity. The Galileon field was shown to encounter problems on
the horizon and within but we went on to show how this singularity
could be eliminated by including linear time dependance in the
manner of \cite{Babichev:2013cya}. Motivated by the differing
approaches of \cite{BBMB} and \cite{Babichev:2013cya} we then went
on to consider a slightly different bi-scalar-tensor theory which invoked some of the ideas of
both approaches. In this way we eventually succeeded in finding two asymptotically flat
hairy scalar tensor black holes where both  scalars are regular
and furthermore the one associated to the BBMB action carries
primary scalar hair.

The construction we have put together in the last section of this
paper can be extended to other terms appearing in the Horndeski
Lagrangian and this will allow to find differing black hole
solutions of the Horndeski theory in a closed form. It would also
be interesting to understand the role of the  integration constant
$q$ associated to time dependence of the scalar field. Is it a
real charge? After all it is not the Galileon $\Psi$ that appears
in the action, it is the derivative. In this sense it is also not
clear that one should worry about regularity of the field $\Psi$
itself, one should maybe worry more about its derivative in the
same way one does for the Maxwell field strength rather than the
potential. A careful study of the underlying thermodynamic
properties would also be interesting. This would most likely shed
some light in the question of relevant and non relevant charges of
the solutions and how they make compare to possibly GR
solutions-if one can find a common thermodynamic bath.

{\bf Acknowledgments}:~~We thank Jorge Zanelli for usefull
discussions.  Additionally we thank the participants of the
"Meeting on the Horizon conference", held in Valparaiso for their
comments during the presentation of various results of this
project. C.C. is delighted to thank Eugeny Babichev and Mokhtar Hassaine for discussions.
M.T. thanks the Laboratoire de Physique Th\'eorique
(LPT), Univ. Paris-Sud in Orray and the National Technical
University of Athens (N.T.U.A.) for hospitality during various
stages of this work. T.K. and E.P. thank the Centro de Estudios
Cientificos (CECs) for hospitality during the initial and final
stages of this work. MT was funded by the FONDECYT Grant No.
3120143. The Centro de Estudios Cientificos (CECs) is funded by
the Chilean Government through the Centers of Excellence Base
Financing Program of Conicyt. E.P is supported partially by
ARISTEIA II action of the operational programme education and long
life learning which is co-funded by the European Union (European
Social Fund) and National Resources. T.K. was funded by the
FONDECYT Grant No. 3140261.

\begin{appendix}

\section{Variations}

Varying (\ref{extended action}) with respect to $\phi$ we get:

 \bea &&{\mathcal E}_{\phi}=
\eta \left(\square \phi-\frac{1}{6}R \phi \right)\nn \\
&&\quad \quad +\g \Big[\nabla_{\n}\left(\nabla^{\n}\Psi\nabla_{\m}\phi\nabla^{\m}\Psi \right)-\frac{1}{2}\nabla^{\a}\left(\nabla_{\a}\phi\nabla_{\m}\Psi\nabla^{\m}\Psi \right)+\frac{1}{3}\nabla^{\a}\left(\nabla_{\a}\phi\nabla_{\m}\Psi\nabla^{\m}\Psi \right)-\frac{1}{6}\square \phi \nabla_{\m}\Psi\nabla^{\m}\Psi\nn \\
&&\quad \quad \quad-\frac{1}{3}\nabla_{\m}\left(\nabla^{\m}\Psi\nabla_{\n}\phi\nabla^{\n}\Psi \right)+\frac{1}{6}\nabla_{\m}\nabla_{\n}\phi\nabla^{\m}\Psi\nabla^{\n}\Psi\nn \\
&&\quad \quad \quad-\frac{1}{6}\phi\, G_{\m\n}\nabla^{\m}\Psi\nabla^{\n}\Psi+\frac{1}{6}\nabla_{\n}\left(\nabla^{\n}\Psi\nabla_{\m}\phi\nabla^{\m}\Psi \right)+\frac{1}{6}\square\Psi\,\nabla_{\n}\phi\nabla^{\n}\Psi+\frac{1}{6}\phi\left( \square\Psi\right)^{2}\nn \\
&&\quad \quad \quad+\frac{1}{6}\nabla_{\n}\phi\nabla^{\m}\Psi\nabla_{\m}\nabla^{\n}\Psi+\frac{1}{6}\phi\nabla_{\m}\nabla_{\n}\Psi\nabla^{\m}\nabla^{\n}\Psi-\frac{1}{6}\nabla^{\a}\left(\nabla_{\a}\phi\nabla_{\m}\Psi\nabla^{\m}\Psi \right)-\frac{1}{3}\nabla_{\m}\phi\nabla_{\n}\Psi\nabla^{\m}\nabla^{\n}\Psi\nn \\
&&\quad \quad \quad
-\frac{1}{3}\phi\,\nabla_{\m}\nabla_{\n}\Psi\nabla^{\m}\nabla^{\n}\Psi-\frac{1}{6}\phi\,R_{\m\n}\nabla^{\m}\Psi\nabla^{\n}\Psi
\Big]=0~, \eea where when collecting similar terms, the above
expression is written as
\bea &&{\mathcal E}_{\phi}=
\eta \left(\square \phi-\frac{1}{6}R \phi \right)\nn \\
&&\quad \quad +\g \Big[\frac{5}{6}\nabla_{\n}\left(\nabla^{\n}\Psi\nabla_{\m}\phi\nabla^{\m}\Psi \right)-\frac{1}{3}\nabla^{\a}\left(\nabla_{\a}\phi\nabla_{\m}\Psi\nabla^{\m}\Psi \right)-\frac{1}{6}\square \phi \nabla_{\m}\Psi\nabla^{\m}\Psi+\frac{1}{6}\nabla_{\m}\nabla_{\n}\phi\nabla^{\m}\Psi\nabla^{\n}\Psi\nn \\
&&\quad \quad \quad-\frac{1}{6}\phi\, G_{\m\n}\nabla^{\m}\Psi\nabla^{\n}\Psi+\frac{1}{6}\square\Psi\,\nabla_{\n}\phi\nabla^{\n}\Psi+\frac{1}{6}\phi\left( \square\Psi\right)^{2}-\frac{1}{6}\nabla_{\n}\phi\nabla^{\m}\Psi\nabla_{\m}\nabla^{\n}\Psi\nn \\
&&\quad \quad \quad
-\frac{1}{6}\phi\,\nabla_{\m}\nabla_{\n}\Psi\nabla^{\m}\nabla^{\n}\Psi-\frac{1}{6}\phi\,R_{\m\n}\nabla^{\m}\Psi\nabla^{\n}\Psi
\Big]=0~. \eea
The metric field equations are \bea
&&{\mathcal H}_{\m\n}=\z G_{\m\n}-\eta \left( \frac{1}{2}\nabla_{\m}\phi\nabla_{\n}\phi-\frac{1}{4}g_{\m\n}\nabla_{\a}\phi\nabla^{\a}\phi+\frac{1}{12}\left(g_{\m\n}\square-\nabla_{\m}\nabla_{\n}+G_{\m\n} \right)\phi^{2}\right)\nn\\
&&\qquad  +\b \big( -\frac{1}{2}g_{\m\n}G^{\a\b}\nabla_{\a}\Psi\nabla_{\b}\Psi+2G_{(\m}^{\,\,\,\,\,\la}\nabla_{\n)}\Psi\nabla_{\la}\Psi+\frac{1}{2}R\nabla_{\m}\Psi\nabla_{\n}\Psi-\frac{1}{2}R_{\m\n}\nabla^{\a}\Psi\nabla_{\a}\Psi+\frac{1}{2}g_{\m\n}\big((\square\Psi)^{2}\nn\\
&&\quad \quad \quad-\nabla_{\a}\nabla_{\b}\Psi\nabla^{\a}\nabla^{\b}\Psi-R_{\a\b}\nabla^{\a}\Psi\nabla^{\b}\Psi\big)+\nabla_{\m}\nabla^{\a}\Psi\nabla_{\n}\nabla_{\a}\Psi-\square\Psi\nabla_{\m}\nabla_{\n}\Psi+R_{\m\,\,\,\n}^{\,\,\,\a\,\,\,\,\b}\nabla_{\a}\Psi\nabla_{\b}\Psi\big)\nn\\
&&\quad \quad +\g\big[\frac{1}{4}g_{\m\n}\nabla_{\a}\phi\nabla_{\b}\phi\nabla^{\a}\Psi\nabla^{\b}\Psi -\nabla_{(\m}\phi\nabla_{\n)}\Psi \nabla_{\a}\phi\nabla^{\a}\Psi-\frac{1}{8}g_{\m\n}\nabla_{\a}\phi\nabla^{\a}\phi\,\nabla_{\b}\Psi\nabla^{\b}\Psi\nn\\
&&\quad \quad \quad +\frac{1}{4}\nabla_{\m}\phi\nabla_{\n}\phi\,\nabla_{\b}\Psi\nabla^{\b}\Psi+\frac{1}{4}\nabla_{\a}\phi\nabla^{\a}\phi\,\nabla_{\m}\Psi\nabla_{\n}\Psi+\frac{1}{24}g_{\m\n}\square(\phi^{2})\nabla_{\a}\Psi\nabla^{\a}\Psi-\frac{1}{12}\nabla_{\m}\nabla_{\n}\phi^{2}\,\nabla_{\a}\Psi\nabla^{\a}\Psi\nn\\
&&\quad \quad \quad +\frac{1}{12}\nabla_{(\m}\left(\nabla_{\n)}(\phi^{2})\,\nabla_{\a}\Psi\nabla^{\a}\Psi \right)-\frac{1}{24}g_{\m\n}\nabla^{\b}(\nabla_{\b}\phi^{2}\,\nabla_{\a}\Psi\nabla^{\a}\Psi)-\frac{1}{12}\square(\phi^{2})\nabla_{\m}\Psi\nabla_{\n}\Psi\nn\\
&&\quad \quad \quad -\frac{1}{24}g_{\m\n}\nabla_{\a}\nabla_{\b}\phi^{2}\,\nabla^{\a}\Psi\nabla^{\b}\Psi+\frac{1}{6}\nabla_{\a}\nabla_{(\m}\phi^{2}\,\nabla_{\n)}\Psi\nabla^{\a}\Psi  -\frac{1}{12}\nabla_{\a}(\nabla_{(\m}\phi^{2}\,\nabla_{\n)}\Psi\nabla^{\a}\Psi)+\frac{1}{24}\nabla_{\a}(\nabla^{\a}\phi^{2}\,\nabla_{\m}\Psi\nabla_{\n}\Psi)\nn\\
&&\quad \quad \quad+\frac{1}{24}\,\phi^{2}g_{\m\n}G^{\a\b}\nabla_{\a}\Psi\nabla_{\b}\Psi-\frac{1}{6}\phi^{2}G_{(\m}\,^{\la}\nabla_{\n)}\Psi\nabla_{\la}\Psi-\frac{1}{24}\phi^{2}R\nabla_{\m}\Psi\nabla_{\n}\Psi\nonumber \\
&&\quad \quad \quad+\frac{1}{24}\phi^{2}R_{\m\n}\nabla^{\a}\Psi\nabla_{\a}\Psi+\frac{1}{12}\nabla_{\a}\nabla_{(\m}\big(\phi^{2}\big)\nabla^{\a}\Psi\nabla_{\n)}\Psi+\frac{1}{12}\square\Psi\nabla_{(\m}\big(\phi^{2}\big)\nabla_{\n)}\Psi \nonumber\\
&&\quad \quad \quad-\frac{1}{12}\nabla^{\a}\Psi\nabla_{(\m}\big(\phi^{2}\big)\nabla_{\n)}\nabla_{\a}\Psi-\frac{1}{12}\nabla_{\a}\big(\phi^{2}\big)\nabla_{(\m}\nabla^{\a}\Psi\nabla_{\n)}\Psi+\frac{1}{12}\nabla_{\a}\big(\phi^{2}\big)\nabla^{\a}\Psi\nabla_{\m}\nabla_{\n}\Psi\nonumber \\
&&\quad \quad \quad-\frac{1}{24}\square\big(\phi^{2}\big)\nabla_{\m}\Psi\nabla_{\n}\Psi-\frac{1}{24}\nabla_{\m}\nabla_{\n}\big(\phi^{2}\big)\nabla^{\a}\Psi\nabla_{\a}\Psi-\frac{1}{24}g_{\m\n}\nabla_{\a}\nabla_{\b}\big(\phi^{2}\big)\nabla^{\a}\Psi\nabla^{\b}\Psi\nonumber \\
&&\quad \quad \quad-\frac{1}{12}g_{\m\n}\nabla_{\a}\big(\phi^{2}\big)\nabla^{\a}\Psi\square\Psi+\frac{1}{12}g_{\m\n}\nabla_{\a}\big(\phi^{2}\big)\nabla_{\b}\Psi\nabla^{\a}\nabla^{\b}\Psi+\frac{1}{24}g_{\m\n}\square\big(\phi^{2}\big)\nabla^{\a}\Psi\nabla_{\a}\Psi \nonumber \\
&&\quad \quad \quad-\frac{1}{12}\phi^{2}\Big(\frac{1}{2}g_{\m\n}\big((\square\Psi)^{2}-\nabla_{\a}\nabla_{\b}\Psi\nabla^{\a}\nabla^{\b}\Psi-R_{\a\b}\nabla^{\a}\Psi\nabla^{\b}\Psi\big)+\nabla_{\m}\nabla^{\a}\Psi\nabla_{\n}\nabla_{\a}\Psi\nonumber \\
&&\quad \quad
\quad-\square\Psi\nabla_{\m}\nabla_{\n}\Psi+R_{\m}\,^{\a}\,_{\n}\,^{\b}\nabla_{\a}\Psi\nabla_{\b}\Psi\Big)
\big]=0~. \ee

\end{appendix}


\end{document}